\documentclass{llncs}

\usepackage[a-2u]{pdfx}
\usepackage[utf8]{inputenc}
\usepackage[T1]{fontenc}
\usepackage{xcolor}

\usepackage[breakable, theorems, skins]{tcolorbox}
\usepackage{amsmath}
\tcbset{enhanced}

\usepackage{paralist}

\usepackage{microtype}
\usepackage{listings}
\definecolor{commentcolor}{rgb}{0.5,0.5,0.5}
\definecolor{darkgreen}{rgb}{0.09, 0.45, 0.27}
\definecolor{bgcolor}{rgb}{0.99,0.99,0.99}
\lstdefinelanguage{DeploymentDescriptor}{
    basicstyle=\linespread{0.9}\ttfamily\scriptsize\lsstyle,
    keywords={application,complete,components,statefulness,cardinality,codefile,inputfile,QoSrequirements,type,requests,per,kind,metadata,labels,name,spec,containers,app,template,probes,time,probability,ports,image,containerPort,probe,limits,timingRequirements},
    comment=[l]{\#}
}

\usepackage{fancyhdr}
\fancypagestyle{plain}{%
\fancyhead{}
\fancyfoot{}

\fancyfoot[L]{\scriptsize This is the authors' version of the paper: \textit{D. Khalyeyev, T. Bureš, P. Hnětynka: Towards Characterization of Edge-Cloud Continuum, in Proceedings of ECSA 2022 Tracks and Workshops, pp. 215-230, 2023}.\\ The final authenticated publication is available online at \url{https://doi.org/10.1007/978-3-031-36889-9_16}}
}
\fancypagestyle{empty}{%
\fancyhead{}
\fancyfoot{}

\fancyfoot[L]{\scriptsize This is the authors' version of the paper: \textit{D. Khalyeyev, T. Bureš, P. Hnětynka: Towards Characterization of Edge-Cloud Continuum, in Proceedings of ECSA 2022 Tracks and Workshops, pp. 215-230, 2023}.\\ The final authenticated publication is available online at \url{https://doi.org/10.1007/978-3-031-36889-9_16}}
}

\begin{document}
\clearpage
\pagestyle{plain}

\title{Towards characterization of edge-cloud continuum}

\author{
Danylo Khalyeyev,
Tomas Bureš, 
Petr Hnětynka
}
\institute{Charles University, Czech Republic\\ \email{\{khalyeyev,bures,hnetynka\}@d3s.mff.cuni.cz}
}
\authorrunning{Danylo Khalyeyev et al.}

\maketitle

\begin{abstract}
Internet of Things and cloud computing are two technological paradigms that reached widespread adoption in recent years. These paradigms are complementary: IoT applications often rely on the computational resources of the cloud to process the data generated by IoT devices. The highly distributed nature of IoT applications and the giant amounts of data involved led to significant parts of computation being moved from the centralized cloud to the edge of the network. This gave rise to new hybrid paradigms, such as edge-cloud computing and fog computing. Recent advances in IoT hardware, combined with the continued increase in complexity and variability of the edge-cloud environment, led to an emergence of a new vision of edge-cloud continuum: the next step of integration between the IoT and the cloud, where software components can seamlessly move between the levels of computational hierarchy. However, as this concept is very new, there is still no established view of what exactly it entails. Several views on the future edge-cloud continuum have been proposed, each with its own set of requirements and expected characteristics. In order to move the discussion of this concept forward, these views need to be put into a coherent picture. In this paper, we provide a review and generalization of the existing literature on edge-cloud continuum, point out its expected features, and discuss the challenges that need to be addressed in order to bring about this envisioned environment for the next generation of smart distributed applications.
\end{abstract}


\section{Introduction}
\label{sec:introduction}

Recent years saw rapid development in cloud technologies and a raise of the cloud computing paradigm to widespread adoption~\cite{alam_2020}. This trend is driven by the unique benefits offered by cloud computing, such as high availability and scalability of computational resources~\cite{monroy_2012}. At the same time, the rise of cloud computing has been accompanied by another trend---increasing prevalence of smart end devices (Internet of Things and Cyber-Physical Systems)~\cite{greer2019cyber}. These two trends seem to push the IT ecosystem in opposing directions: while the cloud paradigm favors centralized services, the IoT and CPS paradigms allow for previously unseen levels of decentralization by bringing the ``smart'' properties into the devices that directly interact with the physical world. In order to fully utilize the benefits of vast computational capacities offered by the centralized cloud, and the benefits of ubiquitous availability and versatility of smart end devices, there is a need to provide a unifying paradigm that would connect this diverse ecosystem from the large data centers, through the intermediate computational, network, and storage nodes, all the way to the end devices~\cite{iotfogecc}.

Recognizing the need to address the challenges of increased prevalence, mobility, and network connectivity of end devices, there have been proposed multiple approaches to make cloud computing more distributed, such as edge computing~\cite{edge}, fog computing~\cite{fog}, transparent computing~\cite{zhang2006transparent}, etc. However, these approaches, while providing a remedy to the problems of geographic distribution and mobility, do not propose a coherent framework that could be used to reason about the applications that span multiple layers of this diverse ecosystem. 

In this search for a unifying paradigm, there has been proposed the concept of Edge-Cloud continuum (ECC)~\cite{roundtable} (also referred to as Mobile-Edge-Cloud Continuum~\cite{unifiedmodel} and Device-Edge-Cloud Continuum~\cite{towardsecc}). This concept envisions the cloud-to-end-device chain as a single unified space with flexible boundaries between levels, in which applications coexist in a highly distributed, yet highly interconnected and resource-rich environment.

Since this concept appeared only very recently, there is no common understanding of it, and there are several distinct views on it. Thus, in its current state, edge-cloud continuum is more of a vision than a coherent paradigm. Yet, given that this vision is aimed to provide an insight into the future of a potentially very impactful technology, there is a need in a more systematic and clear-cut understanding of edge-cloud continuum and related concepts.

In this paper, we 
\begin{inparaenum}[(i)]
\item review the existing work on edge-cloud continuum and related concepts,
\item provide a generalized overview of the expected properties and features of ECC based on that review, and 
\item put forward the questions that need to be answered in future research on the topic. 
\end{inparaenum}
With this contribution, we aim to advance the discussion on the future of IoT and edge computing towards a more well-founded vision.

The rest of the paper is structured as follows. Section~\ref{sec:background} provides a brief overview of the developments in cloud computing and IoT that led to the emergence of the idea of ECC. Section~\ref{sec:core} provides an overview of the literature related to ECC and adjacent topics. Section~\ref{sec:common} distills the existing views on ECC into a common picture. Section~\ref{sec:questions} presents a list of open questions about ECC that need to be addressed in further research on the topic, and Section~\ref{sec:conclusion} concludes the paper.

\section{Background}
\label{sec:background}

The idea of edge computing emerged several years ago as a response to the challenges posed by the rapid growth in usage of mobile devices, and the increasing reliance of mobile applications on cloud services~\cite{edge}. The traditional centralized cloud systems could not guarantee a sufficiently low end-to-end latency to accommodate the needs of mobile device users, and as a response to that, the idea to move the computation closer to the users appeared. One of the most commonly adopted ways to do it is using edge servers, also known as cloudlets~\cite{cloudlets}. A cloudlet is a small server that is supposed to be located just one network hop away from the devices it serves. Thus, it is often assumed to be located near the closest gateway (base transceiver station, router, etc.), or no further than immediately above it (from the perspective of network topology). Using the cloudlet architecture helps to improve user experience not only by reducing the latency between a mobile device and a cloud service, but also by masking network outages to some degree, thus providing a more smooth user experience overall~\cite{edge}.

The idea of fog computing~\cite{fog} is closely related to edge computing. While the term edge computing more often appears when talking about mobile devices\cite{shahzadi2017multi}, such as smartphones, fog computing is more often used in the context of Internet of Things\cite{foginiot}. 

There exists certain confusion in the literature around the terms edge computing and fog computing, as different authors see distinctions between these paradigms differently. Sometimes, fog computing is seen as a sub-case of edge computing, one of its multiple implementations~\cite{8016213}. Other authors consider fog computing as an architecture where the whole network is under the control of a single provider, while in the edge computing the network is more fragmented and shared between edge devices that are unaware of each other and have no information about the entire network~\cite{Singh2019}. Yet another view is that edge computing moves only the computation and the data storage closer to the end user, while fog computing is also related to moving other aspects, such as networking, control, and decision making~\cite{YOUSEFPOUR2019289}. Thus, the line between edge and fog computing is blurred and, especially in the context of edge-cloud continuum, it makes little sense to talk separately about edge and fog. Thus, in the rest of the text, we do not distinguish between these terms.

\section{Existing views on edge-cloud continuum}
\label{sec:core}

The first mentions of edge-cloud continuum (ECC) appeared in 2017~\cite{5g}, including under the names cloud continuum~\cite{7876215} and Device-Fog-Cloud continuum\cite{8368537}. Since then, the term and its variations had appeared in hundreds of papers. In most of those papers, the term is mentioned only briefly, without an explanation or a reference. Here, we focus primarily on the works that attempt to define the term or to provide a more general overview of it. The notion of ECC is understood differently in different papers, and different authors focus on different properties. Thus, in this section, we provide a general overview of the features of ECC proposed in the existing literature and the requirements that ECC is expected to fulfill.

\subsection{Main elements of the edge-cloud continuum}

Being the result of a fusion of IoT and cloud, ECC includes all kinds of entities that can be found in IoT and cloud systems.

Generally, the entities within IoT can be categorized into five types~\cite{iottaxonomy}: end devices, gateways, applications, cloud, and administrative monitoring tools. This classification includes both hardware (end devices, gateways) and software (applications, administrative monitoring tools) entities. Applications there are understood as mobile OS applications, while software monitoring tools are primarily web-based tools. This classification features the cloud as a singular entity, without distinguishing the entities within it. This may be sufficient for modeling IoT applications from the perspective of the end devices, but not enough if we want to model a complete ECC ecosystem. In that case, we would need to distinguish components within the cloud, such as containers, virtual machines, and service endpoints. 

The hardware within ECC is highly heterogeneous. It may come in all sizes, from giant data centers to the smallest single-purpose network-connected sensors and microcontrollers. Basically, any device with at least very basic computational capability and network connectivity can be a part of ECC. Software platforms running on that hardware are also very heterogeneous. In general, they can be categorized into~\cite{towardsecc}: device-specific firmware without an OS, real-time OS, language runtime, full OS (e.g. Linux), App OS (e.g. Android Wear), server OS (e.g. Linux + Node.js), and container OS (e.g. Linux + Docker).

Networks also play a large role in ECC, as all devices need to be somehow connected in order to be a part of the same continuum. Networks also consist of both hardware and software elements, since software-defined networking and network function virtualization are considered to be important enabling technologies of ECC~\cite{sdnecc}\cite{sdnslicing}. 

The network technologies and standards used across ECC are also very diverse~\cite{iotfogecc}. This variety is especially high among the wireless networking standards, ranging from the most common technologies like WiFi, Bluetooth, and NFC, to less widely known protocols, such as ZigBee, Z-Wave, MiWi, etc. This variety is partially due to a lack of standardization in this emerging industry (which is expected to improve soon, with new common standards supported by all major IoT platform companies being developed, such as Matter\footnote{\url{https://csa-iot.org/all-solutions/matter/}}). Another reason for this network heterogeneity is the fact that various applications in ECC have different requirements on the networks they run in, most often these are power consumption, latency, and throughput~\cite{iotfogecc}.

Edge-cloud is often discussed in the context of 5G~\cite{5g} and the coming 6G technologies~\cite{6gedge}. These technologies can provide much higher network throughput, but the area served by one base transceiver station is smaller than in the previous cellular networks~\cite{ecloud}. For ECC this means that each cloudlet deployed in such a network would be able to serve the devices in a smaller geographical area. This means that in order to achieve the ``near-zero'' latency that 6G networks are expected to provide to mobile users, more frequent service handovers will be required, thus the need for dynamic adaptation of workload placement will increase (see section~\ref{sec:placement}).

In order to categorize the spectrum of entities involved in ECC, it is usually divided into levels, forming a ``computational hierarchy''. 

In general, almost all authors distinguish at least three levels of computational hierarchy in the edge-cloud continuum. The highest level, almost universally referred to as the cloud, is characterized by abundant, virtually endless resources, geographical remoteness from the end devices, and typically high latency. This level is ideal for placing workloads that do not require low response time but do require large amounts of computational resources. This level can be represented by large private data centers and/or rented sections of public clouds. The workloads running on this level come in the form of virtual machines or (increasingly commonly) containerized microservices.

The middle level is usually referred to as the edge or fog level. It is made of units that are often referred to as fog nodes, edge nodes, or cloudlets~\cite{cloudlets}. These nodes typically provide limited computational, storage, and network capacities that are used by the services serving the end devices in their immediate vicinity. 

Some authors suggest that the future edge-cloud continuum would contain an intermediary level between the fog and the cloud~\cite{ecloud}. This level would consist of groups of geographically adjacent cloudlets formed dynamically according to the task being currently solved. In many ways, such a dynamic, task-oriented grouping is similar to the concept of ensembles that exists in component systems~\cite{zambonelli2011self}. A group like this does not have a central element and relies on dynamic coordination between cloudlets (or individual services running on those cloudlets).

Sometimes, the fog level is modeled as a sequence of N nested layers~\cite{iotfogecc}. A higher layer is assumed to be composed of servers of larger computational capacity that serve smaller servers on a layer below it.

The last level is most commonly referred to as the device level. The terminology may be sometimes confusing since some authors refer to it as edge level~\cite{capillary} (in contrast to the fog level above), or fog level~\cite{kar2022survey} (in contrast to the edge level above). This level contains a large variety of end devices, from the devices running full-fledged operating systems, such as smartphones or laptops, to robots running on real-time operating systems, to network-connected home appliances, to one-purpose smart sensors, such as temperature sensors or cameras. The important distinguishing property of the devices on this level is that they interface directly with people or the physical world. In contrast, all entities on the levels above provide services and resources that are used either by end devices or by people through interfaces available on end devices.

\subsection{Workload placement}
\label{sec:placement}

It is envisioned that the placement of workloads in ECC is dynamic, and is determined by the current conditions in the ECC environment, and by the current needs of each application~\cite{ecloud}. This means that the same software component may run on different hardware nodes, or even on different levels of computational hierarchy, depending on its current environment and resource requirements. There are many factors that can be taken into account when deciding where to place a workload. According to an industry survey, currently, the most influential factors that determine the preferred placement of a workload are security and cost~\cite{roundtable}, however, much more factors can be considered depending on a particular use case. These include platform reliability, application performance, end-to-end latency, government regulations, corporate regulations, software vendor requirements, etc.

Such a large number of potential factors influencing the placement of software components in the continuum, combined with the inherent unpredictability of edge-cloud environment leads to a need to establish mechanisms for adaptive workload placement, which would determine the exact venue of workload execution at runtime. In recent years, there has been a lot of research into QoS-oriented runtime adaptation in the edge-cloud environment~\cite{bellavista2017migration}\cite{mendoncca2019developing}\cite{rossi2020geo}, however, all that research has been focused on the methods of adaptation within one level of the continuum (either cloud or edge/fog). Yet, edge-cloud continuum, as it is envisioned, should be able to run the same software components on different levels of the continuum, depending on the current needs of individual applications and overall circumstances~\cite{towardsecc}\cite{roundtable}\cite{capillary}. This means that the ECC environment has to be able to support fully liquid software~\cite{towardsecc}. 

The term liquid software refers to the kind of software that can move between multiple devices in a seamless manner~\cite{ls}. In the context of ECC, this can mean two things: either seamless migration and handover of software components between the nodes of the same type (which has been already demonstrated~\cite{doan2020follow}), or seamless migration between devices of different types (which is currently only an idea~\cite{towardsecc}). We will refer to software that can perform both kinds of migrations as \textit{fully liquid software}. 

Implementing fully liquid software has two major challenges. The first is that the migration between different devices has to be performed seamlessly, while continuing normal operation, without introducing any data inconsistencies or impairing user experience. The second challenge is that the same kind of software component has to be able to run on different kinds of devices (different hardware platforms, OS types, etc.). Currently, this is possible only by re-implementing the component for each platform individually, which significantly complicates application development and maintenance, and exacerbates the seamless migration problem.

So far, there is no working framework that would allow implementing fully liquid software in ECC, but some ideas have been suggested, such as using Function-as-a-Service principles to encapsulate common functions into cross-level executable tasks~\cite{pilotedge}\cite{faasls}, and implementing isomorphic software~\cite{isomorphicsw}. The idea of isomorphic software is particularly interesting, since, if implemented, it would allow the same code to run on all levels of ECC, in exactly the same version. This idea requires an underlying platform that would be able to run not only in the cloud and edge/fog but also on IoT devices that are not operated by traditional OSs. The authors of the idea suggest that given the progress in IoT device hardware, eventually, they could run some form of lightweight containers that would encapsulate exactly the same code as the one running on the server side.

\subsection{Artificial intelligence in edge-cloud continuum}
\label{sec:ai-ecc}

One of the key characteristics of ECC is the presence of edge intelligence. The term edge intelligence refers to a practice of locating AI-enabled components, in particular ML models, in the edge and device levels of ECC. The main purpose of this is to provide smart on-the-spot analysis of the data generated by the end devices. The amount of these data may be so large that transferring them over the network all the way to the cloud might be quite expensive and lead to high network bandwidth consumption. At the same time, such network transfers may incur a high latency cost on the results of that analysis.

There are two primary computational processes involved in machine learning: training and inference. These processes can potentially be located at any level across the device-to-cloud hierarchy or even on a combination of levels. Various authors distinguish six~\cite{intelligence} or seven~\cite{6gedge} levels of ML placement within ECC. These levels include in-cloud training and inference, cloud training with cloud-edge co-inference, cloud-edge co-training and co-inference, in-edge training and inference, on-device training and inference, and intermediary options between these. 

Apart from edge training and edge inference, edge intelligence also includes two other processes: edge caching and edge offloading~\cite{9596610}. \textit{Edge caching} involves storing the data generated on devices with little storage and processing capability on more powerful nodes, such as vehicle-located cloudlets or edge servers. The stored data are then used as inputs to both training and inference processes. Apart from storing raw data (e.g. video feeds from cameras), edge caching is also used for storing the results of inference operations (e.g. results of object recognition on video streams). This is done in order to reduce the usage of computational resources on inference and to make those results available to edge devices that lack their own inferencing capability.

The term \textit{edge offloading} refers to the practice of transferring the processes of edge training, edge inference, and edge caching to another hardware node, typically when additional computational resources are needed. Edge offloading is particularly important since it makes it possible to utilize the full spectrum of resources of ECC. Offloading may be done either to an upper level of computational hierarchy (device-to-edge or device-to-cloud offloading) or to another device on the same level (device-to-device offloading)~\cite{9596610}. Ideally, edge offloading should be implemented in a fully transparent way, in which case it can be considered to be a type of liquid software.

\begin{figure}[h!]
  \centering
  \includegraphics[width=\linewidth]{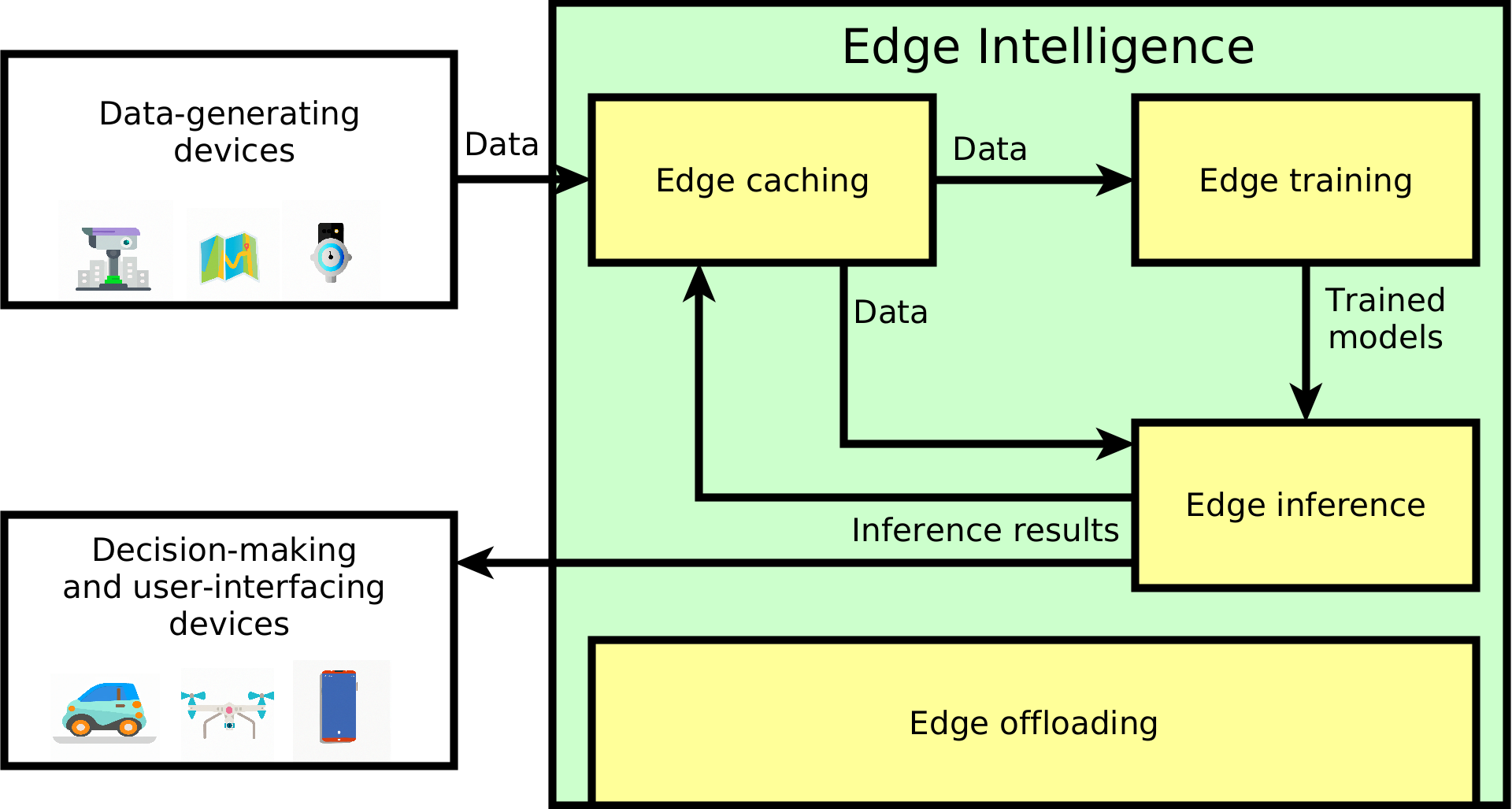}
  \caption{The processes making up edge intelligence and the devices interfacing with it. Arrows represent data flow. Edge offloading underpins the other edge intelligence processes, providing them with additional computational resources when needed.}
  \label{fig:ei}
\end{figure}

\autoref{fig:ei} shows the relationship between the processes involved in edge intelligence and the devices that interact with it.

Edge intelligence, just like many other features of ECC, is in large part enabled by the recent progress in the end device hardware. In recent years, there has been a boom of new kinds of dedicated Edge AI application-specific integrated circuits designed specifically to run on the edge, such as vision processing units (e.g., Intel Movidius VPUs\footnote{\url{https://www.intel.com/content/www/us/en/products/details/processors/movidius-vpu.html}}), tensor processing units (e.g., Google Edge TPU\footnote{\url{https://cloud.google.com/edge-tpu}}), and neural processing units (e.g., Arm  Ethos U55 Machine Learning Processor\footnote{\url{https://www.arm.com/products/silicon-ip-cpu/ethos/ethos-u55}}).

Another reason for employing edge intelligence is privacy and security concerns~\cite{plastiras2018edge}. When the data generated on edge is sufficiently sensitive, stakeholders might not be willing to share it with a third party, such as a cloud service provider. Edge analytics, in this case, becomes a good alternative to establishing a private on-premises cluster to deal with the sensitive part of the data only, as it cuts the cost and the complexity of the system significantly. Some edge AI hardware producers, such as Coral\footnote{\url{https://www.coral.ai/}}, highlight data privacy as one of the most important features of their solutions, along with the ability to work offline, in the areas where connectivity might be limited.

\section{Common properties}
\label{sec:common}

Based on the analysis of the existing literature related to ECC, we can make some conclusions about the properties it is expected to have. In this section, we synthesize the existing views on the edge-cloud continuum and distill their common properties. Below, we focus on each of these properties individually.

\textbf{Hierarchical structure}
The concept of edge-cloud continuum comes from the idea that there is a computational continuum of devices of various capabilities existing between the cloud, with its virtually endless resources, and the substantially resource-constrained end devices. In between these two extremes, there may exist a whole array of nodes of different capabilities. These include smaller servers that can serve a limited number of (typically closely located) clients, smartphones and other consumer devices that, while having limited resources, still can host multiple applications with open-ended functionality, and smart devices, such as wearables, smart home appliances, smart urban sensors, etc. that are often designed to carry out a limited number of simple tasks only. A three-tiered structure is often assumed (device-fog-cloud), but the actual topology may be more complex. The number of levels does not even have to be the same across the whole ECC, with different applications utilizing the computational continuum in different ways. The amount of memory, storage, network capacity, and computational power, as well as connectivity to other devices in the network, is higher on the higher levels of the hierarchy.

Thus, the hierarchical structure of ECC is, first of all, a fact about the hardware of the devices involved in the ECC environment. It is important to understand, though, that given the continued increase in hardware heterogeneity, some devices may not have a properly defined ``place in the hierarchy''. The important feature of this architecture is that a potentially resource-constrained device may rely on the resources available ``higher up'' the hierarchy when the need arises. Thus, the same device may theoretically be considered to be on different levels of the hierarchy with respect to different kinds of resources.

\textbf{Cross-level situation-aware cooperation between components}
The software written for ECC not only has to run in a highly heterogeneous environment, but also to be able to adapt to very variable and unpredictable conditions. The diverse ECC environment offers unparalleled opportunities for self-adaptation, but seizing these opportunities would require adopting less traditional software development paradigms. 

In particular, this may mean developing software consisting of components with a substantial degree of autonomy. These components then might be able to form dynamic task-oriented groups (ensembles), either in the course of their regular operation or as a way to resolve some ``critical situations''. These groups can be formed either between the components of the same kind or between different kinds of components potentially spanning multiple levels of ECC. 

\textbf{Fully liquid software}
Another mechanism that will make the ECC applications more adaptable and resilient towards unpredictable changes is software liquidity, which may be expressed as either ability of software components to migrate between the levels of the computational continuum, or their ability to offload certain functions and responsibilities to the higher levels of the continuum.

A software component that belongs to an ECC application may run on different levels of the computational continuum, depending on current circumstances, like network conditions, current workload, or resource availability. In order to ensure that, the component must be able to run in different environments (different hardware platforms, operating systems, etc.). In addition, seamless handovers during such migrations must be ensured. 

Software liquidity may also refer to the ability of software components to migrate between nodes of the same type in different locations. This kind of migration might help to ensure low latency and reduce network usage (e.g., if it is done to follow client mobility), but is not enough to address resource constraints and environmental variability, for which a solution like isomorphic software is needed.

\textbf{Edge intelligence}
Rapid development and proliferation of AI technologies, particularly deep learning, is another contemporary trend that accompanies the current shift towards edge computing. The convergence of these trends is likely to continue, as AI can potentially provide means of achieving some of the above-mentioned ``smart environment'' properties of ECC. In particular, dynamic situation-aware cooperation in ECC can be achieved by enhancing software components with properties of agents with cognitive capabilities, such as the ability to learn and proactively determine the course of action with respect to self-adaptation.

Additionally, the highly distributed architecture of ECC provides new opportunities for machine learning applications. The enormous amounts of data generated by end devices can be processed on the spot (by means of device-edge-cloud co-training and co-inference) without overloading the networks with excessive data transfers. Thus, we can expect to see larger usage of federated learning and device-edge-cloud co-training and co-inference.

\subsection{Reference example}

\label{sec:example}

\begin{figure}[h!]
  \centering
  \includegraphics[width=\linewidth]{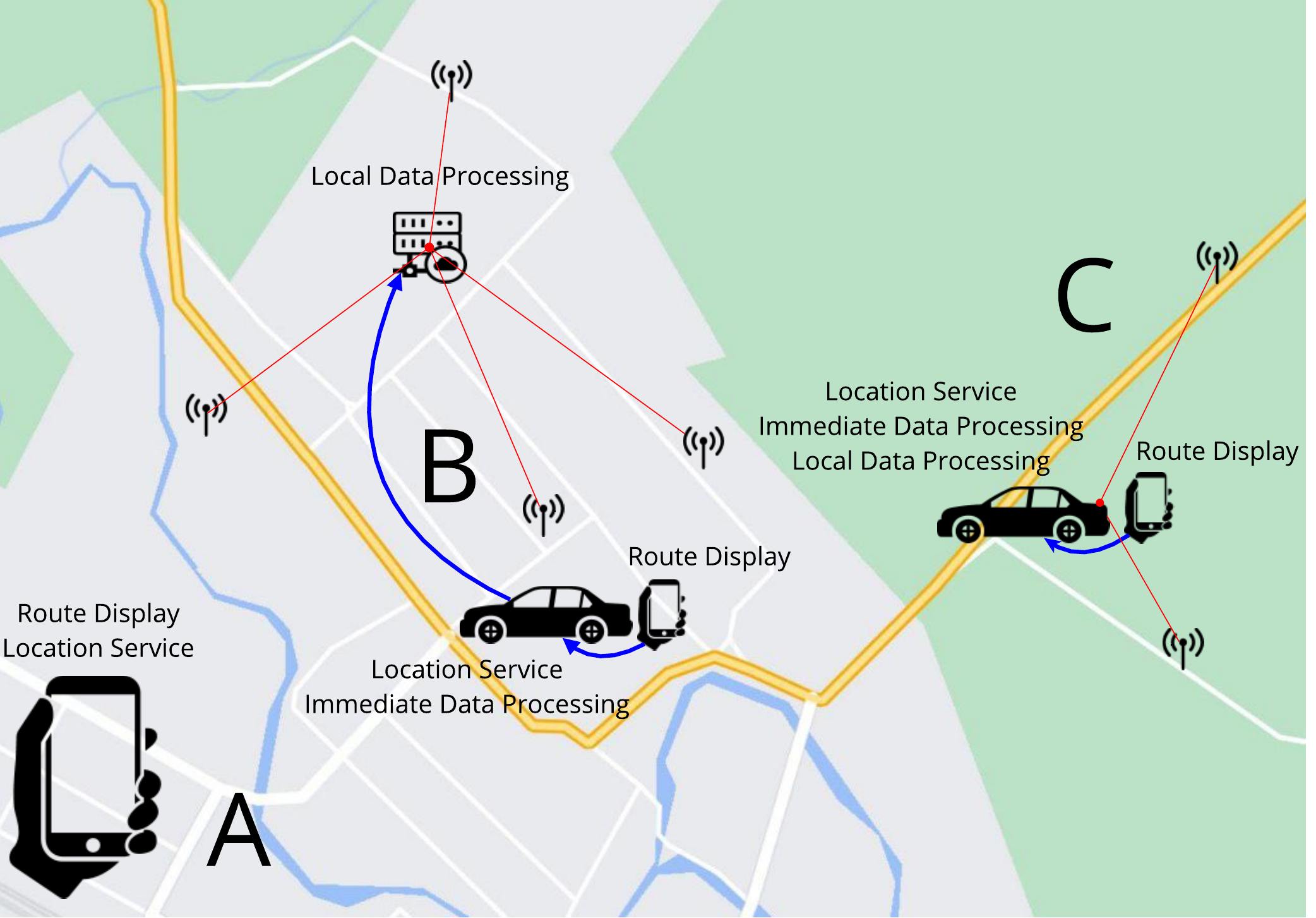}
  \caption{Illustration of the reference example. Case A: a user device relies on its own location service. Case B: the location service component is offloaded to the car, while the car itself relies on local edge servers to process the data from sensors in the neighborhood. Case C: the car runs the local data processing on its own hardware.}
  \label{fig:map}
\end{figure}

Having discussed the trademark characteristics of ECC, we can now construct an example of an application that would demonstrate these characteristics and showcase the benefits that such an environment could provide. This example is centered around smart mobility, real-time urban situation awareness, and route planning. Figure~\ref{fig:map} provides an illustration of the example.

In this example, an end user wants to find a route from its current location to a particular point within a city.  The user is equipped with an internet-connected smartphone that has a route planning application installed. The application uses the cloud services located in a local edge data center in order to carry out the route computation. In order to get to its destination, the user may use one of the smart vehicles (SV) available through a car sharing application. Once the user enters an SV, the route and travel duration computation is offloaded to a cloudlet located in the vehicle itself. At this point, the user's smartphone does not use its own GPS to determine its location and does not make direct requests regarding its route to the cloud, instead relying on more accurate data provided by the SV. 

While in the urban environment, the SV relies on the cloud services (running in the local edge nodes) that gather and process the data from millions of sensors, cameras, and other smart vehicles located throughout the city. Thus, in the urban environment, where major computational nodes are locally available, the SV directly processes only the data from its immediate environment. All the higher-level calculations regarding the route and traffic conditions that are based on the data from thousands of sources are offloaded to the local edge servers. Outside of the urban environment, where there are much fewer sources of information available and where the network connection is not as reliable, the SV performs those calculations on its local hardware. 

The centralized parts of the route-planning application (such as authentication and personal data storage) are handled by the cloud services that may not be located anywhere close to the end user.

This example demonstrates all the properties of an integrated edge-cloud continuum: 

\begin{enumerate}
\item Hierarchical organization. Devices of various sizes and computational capabilities (from smart sensors, through smartphones, smart cars, and edge servers, to data centers) are vertically integrated into one seamless ecosystem.
\item Cross-level cooperation. The components of the ecosystem cooperate with each other in a situation-aware way to deliver an experience of a single ``smart space'' to the end user. This allows a smartphone user to rely upon data generated by thousands of remotely located end devices (cameras, sensors, vehicles) in real time.
\item Fully liquid software. Migration of software functions between different devices, both on the same level (migration between different edge servers) and between the levels (function offloading from smartphone to SV and from SV to the cloud).
\item Edge intelligence. The data collected from various sensors on the far edge of the network are being processed locally by edge servers, and are being used for decision-making locally by smart vehicles.
\end{enumerate}

\section{Open questions}
\label{sec:questions}

In this paper, we focused on answering the question of what ECC is and what are its main characteristics. Of course, these are only the first foundational questions that need to be answered in order to define this area of research more properly. Before the more technical questions related to the details of the future ECC implementation can be asked, we need to put forward some questions about its relevance and feasibility. Edge-cloud continuum seems like a promising new paradigm, and some authors believe that it is bound to transform the way we interact with everyday technology~\cite{ecloud}. However, since not all technological visions eventually become a reality, it is necessary to ask how realistic the realization and adoption of ECC principles are. In particular, we think the following questions are important to answer.
 
\textbf{Q1: Will hardware capabilities be sufficient, and how soon?}
We have mentioned several times that one of the main drivers of edge computing that eventually led to the emergence of the vision of edge-cloud continuum was progress in IoT hardware. Some elements of this vision assume continued progress in that area. And while we probably can expect some further incremental quantitative improvements in hardware capabilities, it is not guaranteed that they will be sufficient to provide a platform for the implementation of some envisioned features of ECC. For instance, achieving fully isomorphic software~\cite{towardsecc} would require end devices to be able to host containerized applications, at least in some form. Whether this is possible in principle, especially for devices running real-time operating systems, is not yet clear.

\textbf{Q2: To what degree is it possible to achieve ECC properties with current software approaches?}
As it was mentioned before, achieving some of the envisioned properties of ECC would require advanced self-adaption at runtime. Currently, there exist some examples demonstrating architectural self-adaptation targeted at edge-cloud scenarios~\cite{bellavista2017migration}\cite{mendoncca2019developing}\cite{rossi2020geo}, and none of them demonstrate the properties of ECC. AI is sometimes suggested as a technology that would allow achieving architectural self-adaptation in such a complex environment~\cite{6gedge}; however, it remains to be seen whether it is suitable for the task. There needs to be more research into AI-directed architectural self-adaptation in ECC, as well as into the self-adaptation mechanisms suitable for edge-cloud continuum in general. 

\textbf{Q3: Are there sufficient incentives to adopt the ECC paradigm in industry, and how to create them?}
Even if this paradigm is technologically feasible and beneficial for the end users, its eventual adoption is dependent on the willingness of major players in the industry to move towards it. This question is especially complex in the case of ECC since its adoption requires cooperation between many different parties, such as infrastructure maintainers, mobile operators, cloud service providers, application developers, etc. All of these parties need to see a sufficient benefit in moving towards ECC in order for it to become a reality. Thus, a more comprehensive analysis of economic incentives behind adopting the principles of ECC is needed.

It must also be noted, that the name edge-cloud continuum itself might not survive the transition from theory to practice: in the future, the environment with the properties outlined in this paper may be referred to differently.

\textbf{Q4: What are the representative examples?}
Perhaps one of the best ways to establish the relevance of the ECC paradigm would be to identify realistic real-world examples of applications that will be enabled by ECC that are not possible to implement within the existing paradigms. Yet, most examples used in the relevant literature today are too far from the state of the art to be considered realistic by the industry.

For instance, many ECC use case examples feature autonomous vehicles in urban environments~\cite {ecloud}. However, it is quite possible that the timelines for the adoption of smart urban mobility are quite far away in the future. The technologies that are expected to enable the shift towards edge-cloud continuum either already exist or will arrive in the next couple of years, while smart urban mobility might be more than a decade away~\cite{8809655}. Thus, examples more relevant to the current day are required. Having such examples could significantly speed up the process of ECC adoption and make it more attractive to the industry. Some promising areas in which suitable examples could be found are natural disaster early warning~\cite{tsunamiwarning}, and emergency situation management~\cite{s21092974}.

\section{Conclusion}
\label{sec:conclusion}

In this paper, we have provided a review of the existing literature on edge-cloud continuum and the related concepts, trends, and paradigms. As a result of that review, we have provided a generalized look at the properties that define edge-cloud continuum according to the current state of the discussion. These properties are the presence of a hierarchy of devices with different computational capabilities, dynamic cross-level cooperation between application components, software liquidity, and ubiquitous edge intelligence. We have also raised some important questions that are important for understanding whether this technological vision can become a reality.

The vision of edge-cloud continuum aims to provide insight into how people will interact with smart technology in the near future. It presents a picture of a ``smart space'' in which people will have instantaneous access to a wealth of applications and devices, seamlessly interacting with each other and the physical world. There still are a lot of hurdles to overcome on the path toward ECC, and some of them are yet to be identified. In this work, we have put this vision into more concrete terms, highlighting its current limitations and the areas of future research.

\section*{Acknowledgment}

This work has been partially supported by Charles University institutional funding SVV 260698/2023, partially by the Charles University Grant Agency project 408622, and partially by the European Research Council (ERC) under the European Union’s Horizon 2020 research and innovation programme (grant agreement No 810115).

\bibliographystyle{splncs04}
\bibliography{paper}

\begin{thebibliography}{10}
\providecommand{\url}[1]{\texttt{#1}}
\providecommand{\urlprefix}{URL }
\providecommand{\doi}[1]{https://doi.org/#1}

\bibitem{alam_2020}
Alam, T.: Cloud computing and its role in the information technology. IAIC
  Transactions on Sustainable Digital Innovation (ITSDI)  \textbf{1}(2),
  108--115 (Feb 2020). \doi{10.34306/itsdi.v1i2.103}

\bibitem{ecloud}
Arulraj, J., Chatterjee, A., Daglis, A., Dhekne, A., Ramachandran, U.:
  {eCloud}: A vision for the evolution of the edge-cloud continuum. Computer
  \textbf{54}(5),  24--33 (2021). \doi{10.1109/MC.2021.3059737}

\bibitem{tsunamiwarning}
Balouek-Thomert, D., Renart, E.G., Zamani, A.R., Simonet, A., Parashar, M.:
  Towards a computing continuum: Enabling edge-to-cloud integration for
  data-driven workflows. The International Journal of High Performance
  Computing Applications  \textbf{33}(6),  1159--1174 (2019).
  \doi{10.1177/1094342019877383}

\bibitem{unifiedmodel}
Baresi, L., Mendon\c{c}a, D.F., Garriga, M., Guinea, S., Quattrocchi, G.: A
  unified model for the mobile-edge-cloud continuum. ACM Trans. Internet
  Technol.  \textbf{19}(2) (apr 2019). \doi{10.1145/3226644}

\bibitem{bellavista2017migration}
Bellavista, P., Zanni, A., Solimando, M.: A migration-enhanced edge computing
  support for mobile devices in hostile environments. In: Proceedings of IWCMC
  2017, Valencia, Spain). pp. 957--962. IEEE (2017).
  \doi{10.1109/IWCMC.2017.7986415}

\bibitem{iotfogecc}
Bittencourt, L., Immich, R., Sakellariou, R., Fonseca, N., Madeira, E., Curado,
  M., Villas, L., DaSilva, L., Lee, C., Rana, O.: The internet of things, fog
  and cloud continuum: Integration and challenges. Internet of Things
  \textbf{3-4},  134--155 (2018).
  \doi{https://doi.org/10.1016/j.iot.2018.09.005}

\bibitem{foginiot}
Bonomi, F., Milito, R., Zhu, J., Addepalli, S.: Fog computing and its role in
  the internet of things. In: Proceedings of MCC 2012, Helsinki, Finland. p.
  13–16. ACM (2012). \doi{10.1145/2342509.2342513}

\bibitem{5g}
Carmo, M.S., Jardim, S., Neto, A.V., Aguiar, R., Corujo, D.: Towards fog-based
  slice-defined wlan infrastructures to cope with future 5g use cases. In:
  Proceedings of NCA 2017, Cambridge, MA, USA (2017).
  \doi{10.1109/NCA.2017.8171397}

\bibitem{7876215}
Carrega, A., Repetto, M.: A network-centric architecture for building the cloud
  continuum. In: Proceedings of ICNC 2017, Silicon Valley, CA, USA. pp.
  701--705 (2017). \doi{10.1109/ICCNC.2017.7876215}

\bibitem{sdnecc}
Dai, M., Su, Z., Li, R., Yu, S.: A software-defined-networking-enabled approach
  for edge-cloud computing in the internet of things. IEEE Network
  \textbf{35}(5),  66--73 (2021). \doi{10.1109/MNET.101.2100052}

\bibitem{fog}
{Dastjerdi}, A.V., {Buyya}, R.: Fog computing: Helping the internet of things
  realize its potential. Computer  \textbf{49}(8),  112--116 (2016).
  \doi{10.1109/MC.2016.245}

\bibitem{doan2020follow}
Doan, T.V., Fan, Z., Nguyen, G.T., Salah, H., You, D., Fitzek, F.H.: Follow me,
  if you can: A framework for seamless migration in mobile edge cloud. In:
  Proceedings of {IEEE} {INFOCOM} 2020 {Workshops}, Toronto, ON, Canada. pp.
  1178--1183. IEEE (2020). \doi{10.1109/INFOCOMWKSHPS50562.2020.9162992}

\bibitem{8016213}
Dolui, K., Datta, S.K.: Comparison of edge computing implementations: Fog
  computing, cloudlet and mobile edge computing. In: Proceedigns of {GIoTS}
  2017, Geneva, Switzerland (2017). \doi{10.1109/GIOTS.2017.8016213}

\bibitem{ls}
Gallidabino, A., Pautasso, C., Mikkonen, T., Systa, K., Voutilainen, J.P.,
  Taivalsaari, A.: Architecting liquid software. Journal of Web Engineering
  \textbf{16}(5–6),  433--470 (2017)

\bibitem{greer2019cyber}
Greer, C., Burns, M., Wollman, D., Griffor, E., et~al.: Cyber-physical systems
  and internet of things. Special Publication (NIST SP) - 1900-202 (2019).
  \doi{10.6028/NIST.SP.1900-202}

\bibitem{8809655}
Jameel, F., Chang, Z., Huang, J., Ristaniemi, T.: Internet of autonomous
  vehicles: Architecture, features, and socio-technological challenges. IEEE
  Wireless Communications  \textbf{26}(4),  21--29 (2019).
  \doi{10.1109/MWC.2019.1800522}

\bibitem{kar2022survey}
Kar, B., Yahya, W., Lin, Y.D., Ali, A.: A survey on offloading in federated
  cloud-edge-fog systems with traditional optimization and machine learning.
  arXiv preprint arXiv:2202.10628  (2022). \doi{10.48550/arXiv.2202.10628}

\bibitem{pilotedge}
Luckow, A., Rattan, K., Jha, S.: Pilot-edge: Distributed resource management
  along the edge-to-cloud continuum. In: Proceedings of IPDPSW 2021, Portland,
  OR, USA. pp. 874--878 (2021). \doi{10.1109/IPDPSW52791.2021.00130}

\bibitem{8368537}
Martin, B.A., Michaud, F., Banks, D., Mosenia, A., Zolfonoon, R., Irwan, S.,
  Schrecker, S., Zao, J.K.: Openfog security requirements and approaches. In:
  Proceedings of IEEE Fog World Congress (FWC), Santa Clara, CA, USA (2017).
  \doi{10.1109/FWC.2017.8368537}

\bibitem{s21092974}
Masip-Bruin, X., Marín-Tordera, E., Sánchez-López, S., Garcia, J., Jukan,
  A., Juan~Ferrer, A., Queralt, A., Salis, A., Bartoli, A., Cankar, M.,
  Cordeiro, C., Jensen, J., Kennedy, J.: Managing the cloud continuum: Lessons
  learnt from a real fog-to-cloud deployment. Sensors  \textbf{21}(9) (2021).
  \doi{10.3390/s21092974}, \url{https://www.mdpi.com/1424-8220/21/9/2974}

\bibitem{mendoncca2019developing}
Mendon{\c{c}}a, N.C., Jamshidi, P., Garlan, D., Pahl, C.: Developing
  self-adaptive microservice systems: Challenges and directions. IEEE Software
  \textbf{38}(2),  70--79 (2019). \doi{10.1109/MS.2019.2955937}

\bibitem{isomorphicsw}
Mikkonen, T., Pautasso, C., Taivalsaari, A.: Isomorphic internet of things
  architectures with web technologies. Computer  \textbf{54}(7),  69--78
  (2021). \doi{10.1109/MC.2021.3074258}

\bibitem{roundtable}
Milojicic, D.: The edge-to-cloud continuum. Computer  \textbf{53}(11),  16--25
  (nov 2020). \doi{10.1109/MC.2020.3007297}

\bibitem{6gedge}
Peltonen, E., Bennis, M., Capobianco, M., Debbah, m., Ding, A.,
  Gil-Castiñeira, F., Jurmu, M., Karvonen, T., Kelanti, M., Kliks, A.,
  Leppänen, T., Lovén, L., Mikkonen, T., Rao, A., Samarakoon, S., Seppänen,
  K., Sroka, P., Tarkoma, S., Yang, T.: 6g white paper on edge intelligence.
  CoRR  \textbf{abs/2004.14850} (2020). \doi{10.48550/arXiv.2004.14850}

\bibitem{plastiras2018edge}
Plastiras, G., Terzi, M., Kyrkou, C., Theocharidcs, T.: Edge intelligence:
  Challenges and opportunities of near-sensor machine learning applications.
  In: Proceedings of ASAP 2018, Milan, Italy (2018).
  \doi{10.1109/ASAP.2018.8445118}

\bibitem{monroy_2012}
Rodríguez-Monroy, C., Arias, C., Núñez~Guerrero, Y.: The new cloud computing
  paradigm: the way to {IT} seen as a utility. Latin American and Caribbean
  Journal of Engineering Education  \textbf{6},  24--31 (dec 2012)

\bibitem{rossi2020geo}
Rossi, F., Cardellini, V., Presti, F.L., Nardelli, M.: Geo-distributed
  efficient deployment of containers with kubernetes. Computer Communications
  \textbf{159},  161--174 (2020). \doi{10.1016/j.comcom.2020.04.061}

\bibitem{edge}
{Satyanarayanan}, M.: The emergence of edge computing. Computer
  \textbf{50}(1),  30--39 (2017). \doi{10.1109/MC.2017.9}

\bibitem{cloudlets}
{Satyanarayanan}, M., {Bahl}, P., {Caceres}, R., {Davies}, N.: The case for
  vm-based cloudlets in mobile computing. IEEE Pervasive Computing
  \textbf{8}(4),  14--23 (2009). \doi{10.1109/MPRV.2009.82}

\bibitem{sdnslicing}
Shah, S.D.A., Gregory, M.A., Li, S.: Cloud-native network slicing using
  software defined networking based multi-access edge computing: A survey. IEEE
  Access  \textbf{9},  10903--10924 (2021). \doi{10.1109/ACCESS.2021.3050155}

\bibitem{shahzadi2017multi}
Shahzadi, S., Iqbal, M., Dagiuklas, T., Qayyum, Z.U.: Multi-access edge
  computing: open issues, challenges and future perspectives. Journal of Cloud
  Computing  \textbf{6}(1),  1--13 (2017). \doi{10.1186/s13677-017-0097-9}

\bibitem{Singh2019}
Singh, S.P., Nayyar, A., Kumar, R., Sharma, A.: Fog computing: from
  architecture to edge computing and big data processing. The Journal of
  Supercomputing  \textbf{75}(4),  2070--2105 (Apr 2019).
  \doi{10.1007/s11227-018-2701-2}

\bibitem{faasls}
Spillner, J.: Self-balancing architectures based on liquid functions across
  computing continuums. In: Proceedings of UCC '21, Leicester, UK (2021).
  \doi{10.1145/3492323.3495589}

\bibitem{capillary}
Taherizadeh, S., Stankovski, V., Grobelnik, M.: A capillary computing
  architecture for dynamic internet of things: Orchestration of microservices
  from edge devices to fog and cloud providers. Sensors  \textbf{18}(9), ~2938
  (2018). \doi{10.3390/s18092938}

\bibitem{iottaxonomy}
Taivalsaari, A., Mikkonen, T.: A roadmap to the programmable world: Software
  challenges in the {IoT} era. IEEE Software  \textbf{34}(1),  72--80 (2017).
  \doi{10.1109/MS.2017.26}

\bibitem{towardsecc}
Taivalsaari, A., Mikkonen, T., Pautasso, C.: Towards seamless iot
  device-edge-cloud continuum:. In: Proceedings of ICWE 2021 Workshops,
  Biarritz, France. pp. 82--98 (2022). \doi{10.1007/978-3-030-92231-3\_8}

\bibitem{9596610}
Xu, D., Li, T., Li, Y., Su, X., Tarkoma, S., Jiang, T., Crowcroft, J., Hui, P.:
  Edge intelligence: Empowering intelligence to the edge of network.
  Proceedings of the IEEE  \textbf{109}(11),  1778--1837 (2021).
  \doi{10.1109/JPROC.2021.3119950}

\bibitem{YOUSEFPOUR2019289}
Yousefpour, A., Fung, C., Nguyen, T., Kadiyala, K., Jalali, F., Niakanlahiji,
  A., Kong, J., Jue, J.P.: All one needs to know about fog computing and
  related edge computing paradigms: A complete survey. Journal of Systems
  Architecture  \textbf{98},  289--330 (2019).
  \doi{https://doi.org/10.1016/j.sysarc.2019.02.009}

\bibitem{zambonelli2011self}
Zambonelli, F., Bicocchi, N., Cabri, G., Leonardi, L., Puviani, M.: On
  self-adaptation, self-expression, and self-awareness in autonomic service
  component ensembles. In: Proceedings of SASOW 2011, Ann Arbor, MI, USA. pp.
  108--113 (2011). \doi{10.1109/SASOW.2011.24}

\bibitem{zhang2006transparent}
Zhang, Y., Zhou, Y.: Transparent computing: A new paradigm for pervasive
  computing. In: Proceedings of UIC 2006, Wuhan, China (2006).
  \doi{10.1007/11833529\_1}

\bibitem{intelligence}
Zhou, Z., Chen, X., Li, E., Zeng, L., Luo, K., Zhang, J.: Edge intelligence:
  Paving the last mile of artificial intelligence with edge computing.
  Proceedings of the IEEE  \textbf{107}(8),  1738--1762 (2019).
  \doi{10.1109/JPROC.2019.2918951}

\end{thebibliography}

\end{document}